\documentclass[twocolumn,prl,a4paper,showpacs,floatfix,preprintnumbers,amsmath,amssymb]{revtex4}

\usepackage{amsmath}
\usepackage{graphicx}

\usepackage{graphicx}
\usepackage{dcolumn}
\usepackage{bm}

\newcommand{\ud}{\mathrm{d}}

\newcommand{\be}{\begin{equation}}
\newcommand{\ee}{\end{equation}}

\begin{document}

\title{Entanglement, Non-linear Dynamics, and the Heisenberg Limit}

\author{Luca Pezz\'e\footnote{New address: Laboratoire Charles Fabry, Institut d'Optique, 
campus Polytechnique, F-91127 Palaiseau cedex, France.} and Augusto Smerzi}
\affiliation{ BEC-CNR-INFM and Dipartimento di Fisica, Universit\`a di Trento, I-38050 Povo, Italy}

\date{\today}
                      
\begin{abstract}
We show that the quantum Fisher information provides a sufficient condition to recognize 
multi-particle entanglement in a $N$ qubit state. 
The same criterion gives a necessary and sufficient condition for sub shot-noise phase sensitivity 
in the estimation of a collective rotation angle $\theta$.
The analysis therefore singles out the class of entangled states 
which are {\it useful} to overcome classical phase sensitivity in metrology and sensors. 
We finally study the creation of useful entangled states  
by the non-linear dynamical evolution of two decoupled 
Bose-Einstein condensates or trapped ions. 
\end{abstract}

\pacs{
03.75.Gg,
03.65.Ud, 
03.67.Bg  
03.75.Dg
}

\maketitle

{\it Introduction}. 
The ability to create and manipulate entangled states of many-particle systems is 
a far-reaching possibility of quantum mechanics.
Several efforts have been devoted, in the last few years, to
exploit entanglement to design new technologies for
secure communication, metrology and fast computation
or to unveil foundational problems of quantum mechanics.
From the experimental point of view,
trapped Bose-Einstein condensates (BECs) \cite{Sorensen_2001, BEC}, 
cold/thermal atoms \cite{coldatoms} and trapped ions \cite{ions} 
are important candidates for the creation of large scale quantum entanglement. 
It is important to emphasize, however,
that not all entangled states are equally {\it useful} 
for developing protocols that outperform classical operations. 
Generally speaking, current measures of entanglement mostly focus on the algebraic 
separability properties of quantum states. 
This notion should be extended for quantum technological applications, 
where it is essential 
to classify entanglement on the basis of some additional physical/algebraic
properties required by the specific task.
These attributes are crucially related with non-separability, 
but are not necessarily possessed by all entangled states.

In this Letter, we develop a general framework to study the interplay between 
entanglement and phase estimation in metrology and quantum sensors \cite{Giovannetti_2004}. 
A quantum state $\hat{\rho}_{\mathrm{inp}}$ must 
necessarily be entangled in order to be useful for estimating a phase shift $\theta$ with a sensitivity 
$\Delta \theta$ beyond the shot-noise, which is the maximum limit attainable with separable states.
Nevertheless not all entangled states can perform better than separable states.
Here we introduce a new criterion, on a generic $\hat{\rho}_{\mathrm{inp}}$,
which is sufficient to recognize multi-particle entanglement
and is necessary and sufficient for sub shot-noise phase estimation sensitivity.
We separate entangled states in two classes on the basis of an additional
geometrical (or kinetic, see below) property in the Hilbert space.
Our analysis uses basic tools of parameter estimation theory and provides a simple
and experimentally measurable condition, Eq.(\ref{ps}), which extends other criteria
discussed in the literature based on the concept
of spin squeezing \cite{Sorensen_2001}. 
We will show, with
an example experimentally achievable with dilute BECs and trapped ions,
how non-linearity can generate a class of states which are entangled, useful for 
sub shot-noise interferometry, but not spin-squeezed. 

A state of $N$ particles in two modes ($N$ qubits) is separable (non-entangled) when it 
can be written as \cite{Sorensen_2001, Peres_1995}
\be \label{rho}
\hat{\rho}_{\mathrm{sep}} = \sum_k p_k \, \hat{\rho}^{(1)}_k \otimes \hat{\rho}^{(2)}_k \otimes ... \otimes \hat{\rho}^{(N)}_k,
\ee
where $p_k>0$, $\sum_{k}p_k=1$ and $\hat{\rho}_k^{(i)}$ is the density matrix for the $i$th particle.
How to recognize entangled states? Let us introduce 
the ``fictitious'' angular momentum operator, 
$\hat{J} = \sum_{l=1}^{N}\hat{\sigma}^{(l)} $
where $\hat{\sigma}^{(l)}$ is a Pauli matrix operating on the $l$th particle.
According to the current literature, if a state $\hat{\rho}_{\mathrm{inp}}$ satisfies the inequality 
\be \label{cz}
\xi^2 \equiv \frac{N (\Delta \hat{J}_{\vec{n}_3})^2}{\langle \hat{J}_{\vec{n}_1} \rangle^2 + \langle \hat{J}_{\vec{n}_2} \rangle^2}<1,
\ee
then is particle-entangled \cite{Sorensen_2001, Toth_2007, Ulam-Orgikh_2001} and spin squeezed \cite{Sorensen_2001, Wineland_1992, Kitagawa_1993}
along the direction $\vec{n}_3$, being
$\vec{n}_1$, $\vec{n}_2$ and $\vec{n}_3$ three mutually orthogonal unit vectors and
$\hat{J}_{\vec{n}_i} = \hat{J} \cdot \vec{n}_i$.

Here we introduce a different sufficient condition for particle-entanglement:
\be \label{ps}
\chi^2 \equiv \frac{N}{F_\mathrm{Q} [\hat{\rho}_{\mathrm{inp}}, \hat{J}_{\vec{n}}]}<1,
\ee
where $F_\mathrm{Q} [\hat{\rho}_{\mathrm{inp}}, \hat{J}_{\vec{n}}] = 4 (\Delta \hat{R})^2$ is the quantum 
Fisher information (QFI) \cite{nota001, Braunstein_1996, Helstrom, Wootters_1981}
and $\vec{n}$ is an arbitrary direction.
The Hermitean operator $\hat{R}$ is the solution of the equation
$\{ \hat{R}, \hat{\rho}_{\mathrm{inp}} \}= i[\hat{J}_{\vec{n}}, \hat{\rho}_{\mathrm{inp}}]$ \cite{nota002}.
It is possible to demonstrate that $\chi^2 \leq \xi^2 $ \cite{unpublished}. 
Therefore, Eq.(\ref{ps}) recognizes a class of states which are entangled, 
$\chi^2<1$ and not spin-squeezed, $\xi^2\geq 1$
as, for instance, the maximally entangled state \cite{Ulam-Orgikh_2001}.
Notice that, for a pure state, $\hat{\rho}_{\mathrm{inp}}=|\psi_{\mathrm{inp}}\rangle \langle \psi_{\mathrm{inp}}|$, 
we have $F_\mathrm{Q} [\hat{\rho}_{\mathrm{inp}}, \hat{J}_{\vec{n}}]=4 (\Delta \hat{J}_{\vec{n}})^2$ \cite{Braunstein_1996} 
and the sufficient condition for multiparticle entanglement,
Eq.(\ref{ps}), assumes the appealing form 
\be \label{ps2}
\chi_{\mathrm{ps}}^2 \equiv \frac{N}{4 (\Delta \hat{J}_{\vec{n}})^2}<1.
\ee
The QFI is naturally related to the problem of phase estimation.
Generally speaking, an interferometer is quantum mechanically described as a collective, linear,
rotation of the input state by an angle $\theta$:
$\hat{\rho}_{\mathrm{out}}(\theta)=e^{i \theta \hat{J}_{\vec{n}}} \hat{\rho}_{\mathrm{inp}} e^{-i \theta \hat{J}_{\vec{n}}}$.
The goal is to estimate $\theta$ with a 
sensitivity overcoming the shot-noise limit $\Delta \theta_{\mathrm{sn}} \equiv 1/\sqrt{N}$.
For instance, in Mach-Zehnder (MZ) interferometry, 
$\theta$ is a relative phase shift among the two arms of the interferometer, 
and the rotation is about the $\vec{n}=\vec{y}$ axis.

For an arbitrary interferometer and phase estimation strategy,
the phase sensitivity is limited by a fundamental bound, the 
Quantum Cramer-Rao (QCR) \cite{Helstrom},
which only depends on the specific choice of the input state,  
\be \label{cr}
\Delta \theta_{\mathrm{QCR}} = \frac{1}{ \sqrt{F_\mathrm{Q} [\hat{\rho}_{\mathrm{inp}}, \hat{J}_{\vec{n}}]} } = \frac{\chi}{\sqrt{N}}.
\ee
A comparison with Eq.(\ref{cr}) reveals that Eq.(\ref{ps}) is not only a sufficient condition 
for particle-entanglement, as already discussed, but also a necessary and sufficient condition 
for sub shot-noise phase estimation. This is a main result of this work: 
$\chi<1$ provides the class of entangled states which are {\it useful} for sub shot-noise sensitivity. 
In other words, chosen $\hat{\rho}_{\mathrm{inp}}$ and $\hat{J} \cdot \vec{n}$, if the corresponding value of the 
QFI is such that $\chi<1$, then the state is entangled and, if used as input of an interferometer 
realizing the unitary transformation $e^{-i \theta \hat{J}_{\vec{n}}}$,
it provides a phase estimation sensitivity higher than any interferometer using classical (separable)
states. On the other hand, the class of entangled states for which $\chi \geq 1$ cannot provide a sensitivity
higher than the classical shot-noise. 
 
The QFI, which links Eqs.(\ref{ps})
and (\ref{cr}), has a simple interpretation as square of a ``statistical speed", 
$\upsilon^2_F \equiv F_\mathrm{Q}[\hat{\rho}_{\mathrm{inp}}, \hat{J}_{\vec{n}}] = (\ud l(\theta)/\ud \theta)^2$. 
This corresponds to the rate of change of the absolute statistical distance $l(\theta)$
among two pure states in the Hilbert space (or in the space of density operators for general mixtures) 
along the path parametrized by $\theta$ \cite{Braunstein_1996, Wootters_1981}.
The absolute statistical distance is the maximum number of distinguishable states along the path 
parametrized by $\theta$, optimized over all possible generalized quantum measurements.
According to Eq.(\ref{ps}), \emph{useful} entanglement corresponds to high speed,  
$|\upsilon_\mathrm{F}| > |\upsilon_{\mathrm{cr}}|$, being $|\upsilon_{\mathrm{cr}}|=\sqrt{N}$ 
a critical velocity that cannot be overcame by separable states Eq.(\ref{rho}).
The maximum speed (strongest entanglement) is $|\upsilon_{\mathrm{max}}| = N$ and therefore
the fundamental (Heisenberg) limit in phase sensitivity is $\Delta \theta_{\mathrm{HL}} = 1/N$.
Physically, this simply means that, under the action of some unitary evolution, useful entangled states evolve 
(become distinguishable) more rapidly than any separable state.

{\it Entanglement.} 
Let us introduce the inequalities
\be \label{inequality}
\frac{1}{M_{2k}(\theta)}  
\bigg( \frac{\ud M_k(\theta)}{\ud \theta} \bigg)^2 \leq F_\mathrm{Q} [\hat{\rho}_{\mathrm{inp}},\hat{J}_{\vec{n}}] \leq 4 (\Delta \hat{J}_{\vec{n}})^2,
\ee
being $M_{k}(\theta) \equiv \mathrm{Tr}[\hat{M}^k \hat{\rho}_{\mathrm{out}}]$, 
and $\hat{M}$ an arbitrary observable \cite{nota004}.
The right-hand side of Eq.(\ref{inequality}) allows us to demonstrate Eq.(\ref{ps})
by showing that $F_\mathrm{Q} [\hat{\rho}_{\mathrm{sep}},\hat{J}_{\vec{n}}] \leq N$  
for any arbitrary unit vector $\vec{n}$ in the pseudo angular momentum space. 
First, notice that, for separable states, 
$\hat{\rho}_k= \hat{\rho}_k^{(1)} \otimes \hat{\rho}_k^{(2)} \otimes ... \otimes \hat{\rho}_k^{(N)}$, 
we have $4 (\Delta \hat{J}_{\vec{n}})^2 = N - 4\sum_{i=1}^N \langle \hat{j}_{\vec{n}}^{(i)} \rangle^2 \leq N$.
Combining this result with Eq.(\ref{inequality}) and the convexity of the QFI \cite{unpublished} 
(i.e. for an arbitrary mixture $\hat{\rho}=\sum_k p_k \hat{\rho}_k$,
$F_\mathrm{Q} [\hat{\rho},\hat{J}_{\vec{n}}] \leq \sum_k p_k F_\mathrm{Q} [\hat{\rho}_k,\hat{J}_{\vec{n}}]$) 
we obtain that $F_\mathrm{Q} [\hat{\rho}_{\mathrm{sep}}, \hat{J}_{\vec{n}}] \leq N$, 
where the equality sign can be saturated only with pure states.
Moreover, since $4(\Delta \hat{J}_{\vec{n}})^2 \leq 4\langle \hat{J}_{\vec{n}}^2 \rangle \leq N^2$, 
we obtain $F_\mathrm{Q} [\hat{\rho}_{\mathrm{inp}},\hat{J}_{\vec{n}}]\leq N^2$.
Then, from Eq.(\ref{cr}), follows that $\Delta \theta_{\mathrm{HL}}$ is the highest possible phase 
sensitivity.

Using the left-hand side of Eq.(\ref{inequality}) 
we now demonstrate that $\chi \leq \xi$ for any arbitrary $\hat{\rho}_{\mathrm{inp}}$. 
We consider, without loss of generality, a direction
$\vec{n} \equiv \vec{n}_2$ such that $\langle \hat{J}_{\vec{n}_2} \rangle=0$.
By choosing $\hat{M}=\hat{J}_{\vec{n}_3} - \langle \hat{J}_{\vec{n}_3} \rangle$ in Eq.(\ref{inequality}),
we obtain that $F_\mathrm{Q} [\hat{\rho}_{\mathrm{inp}},\hat{J}_{\vec{n}}] \geq (\ud M_1 / \ud \theta)^2/M_2 = N/\xi^2$.
Then, Eq.(\ref{ps}) shows that $\chi \leq \xi$:
the class of states satisfying $\chi<1$ is wider 
and includes the class of states defined by Eq.(\ref{cz}).

\begin{figure}[!t]
\begin{center}
\includegraphics[scale=0.5]{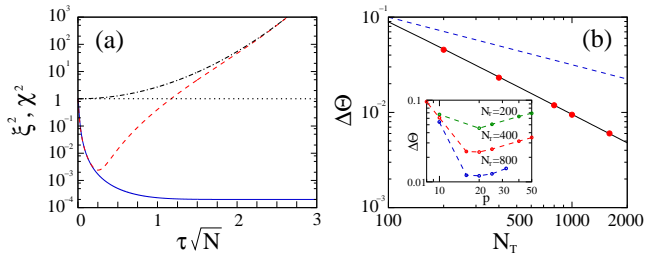}
\end{center}
\caption{\small{
(color online) a) Plot of Eqs.(\ref{xinosq}) (dot-dashed black line), 
(\ref{nosq}) (solid blue line) and (\ref{spsq}) (dashed red line)
as a function of $\tau \sqrt{N}$ (here $N=10^4$).
The states having $\chi^2$, $zeta^2<1$ (i.e. below the horizontal dotted line in the figure)
are useful for quantum interferometry.
b) Phase sensitivity as a function of the total number of particles $N_\mathrm{T}=Np$.
Circles are results of numerical simulations, 
the black line is the Heisenberg limit $\Delta \theta = 8.9/N_\mathrm{T}$, obtained for $p=p_{\mathrm{opt}}$,
and the dotted blue line is the shot-noise $\Delta \theta = 1/\sqrt{N_\mathrm{T}}$.
Inset: $\Delta \theta$ as a function of the number of measurements $p$, 
for fixed values of $N_\mathrm{T}$.
The optimal working point (minimum of each curve) is $p_{\mathrm{opt}}=20$, 
independently from $N_\mathrm{T}$.}} \label{comp} 
\end{figure}

\emph{Non-linear dynamics.}
We now discuss the connection between non-linear dynamics,
entanglement and spin-squeezing.  
We consider a coherent spin state,
$|j, j\rangle_{\vec{n}_1}=\sum_{\mu=-j}^{+j}\frac{1}{2^j}\sqrt{{2j \choose j-\mu}}|j, \mu\rangle_{\vec{n}_3}$ 
\cite{Arecchi_1972, nota13}, with $j=N/2$.
This state is separable ($\chi^2 = 1$) 
and we quest the possibility to strongly entangle the particles 
by the non-linear evolution $e^{-i \tau \hat{J}_{\vec{n}_3}^2}$.
A direct calculation of Eqs.(\ref{cz}) and (\ref{ps}) with $\vec{n} \equiv \vec{n}_2$
(where the expectation values are computed over the state 
$|\psi(\tau)\rangle=e^{- i \tau \hat{J}_{\vec{n}_3}^2} |j, j\rangle_{\vec{n}_1}$)
gives 
\be \label{xinosq}
\xi^2 =  (\cos \tau)^{-2(N-1)},  
\ee
\be \label{nosq}
\chi^2  =  2/\big[(N+1)-(N-1)(\cos 2\tau)^{N-2}\big].
\ee 
Notice that, $\xi^2\geq 1$, while $\chi^2\leq 1$ for all values of $\tau$:
the state $|\psi(\tau) \rangle$ is not spin-squeezed 
but still (usefully) entangled.
A comparison between Eq.(\ref{xinosq}) and Eq.(\ref{nosq}) is presented in Fig.(\ref{comp},a) for $N \gg 1$.
We emphasize two time scales in the dynamical evolution of $\chi^2$: 
for $0<\tau<1/\sqrt{N}$, $\chi^2$ decreases 
from 1 to $2/N$;
for $1/\sqrt{N}\leq \tau \leq \pi/2-1/\sqrt{N}$, it reaches the plateau 
$\chi^2=2/N$.
The dynamics are periodic with period $T=\pi/2$ for even values of $N$ and $T=\pi$ for odd $N$
(in which case $\chi^2=1/N$ at $\tau=\pi/2$).

Kitagawa and Ueda \cite{Kitagawa_1993} have pointed out that 
the non-linear evolution $e^{- i \tau \hat{J}_{\vec{n}_3}^2}$ 
actually creates spin-squeezing, for $\tau \leq 1/\sqrt{N}$, 
along a particular direction.
The maximum squeezing is obtained 
for the state $|\tilde{\psi}(\tau)\rangle=e^{i \delta \hat{J}_{\vec{n}_1}} 
|\psi(\tau)\rangle$, where
$\delta(N,\tau)=\frac{1}{2}\arctan \frac{B}{A}$,
$A=1-(\cos 2\tau)^{N-2}$ and $B=4\sin \tau (\cos \tau)^{N-2}$. 
We have
\be \label{spsq}
\xi^2=\big[4+(N-1)(A-\sqrt{A^2+B^2})\big]/4 (\cos \tau)^{2N-2}.
\ee
Equation (\ref{spsq}), as a function of $\tau$, is shown in Fig.(\ref{comp},a) \cite{nota003}.
We have $\xi^2<1$ for $ 0 < \tau \leq 1.15/\sqrt{N}$ and 
the minimum, $\xi^2_{\mathrm{min}}= 1/N^{2/3}$, is reached at $\tau = 1.2/N^{2/3}$.
For $1/\sqrt{N} \lesssim \tau \geq \pi/2$, $\xi^2 > 1$ and it
converges to Eq.(\ref{xinosq}), which eventually diverges at $\tau=\pi/2$.

{\it Heisenberg Limit.} 
So far we have demonstrated that the non-linear evolution of a coherent spin 
state creates particle entanglement useful for sub shot-noise sensitivity.
This protocol has advantages
when compared to the spin-squeezing approach discussed in \cite{Kitagawa_1993} 
for improving the phase sensitivity of a MZ \cite{Poulsen_2002}.
While spin-squeezing is created only for
short times, $\tau \lesssim 1/\sqrt{N}$, and along a direction $\delta(N,\tau)$ which strongly depends on $\tau$ and $N$, 
our scheme does not require any additional 
rotation of the initial state, is fairly independent on the evolution time and reaches the 
Heisenberg limit \cite{nota18}, $\Delta \theta_{\mathrm{HL}} = 1/N$, for times for $\tau \gtrsim 1/\sqrt{N}$.
Here we apply these results to a realistic BEC experimental setup.
The coherent spin state can be created by splitting an initial condensate in two modes with 
the ramping of a potential barrier
or by quickly transferring half of the particles from an initial 
condensate to two different hyperfine levels with a $\pi/2$ Bragg pulse.
The non-linear evolution, $e^{-i \tau \hat{J}_{z}^2}$, 
where $\tau=E_c t$, $E_c$ is the charging energy and $t$ is the evolution time, 
is naturally provided by particle-particle interaction \cite{nota20}.
The non-linear dynamics of an initial separable state has been 
also recently experimentally demonstrated with trapped ions \cite{ions}.
Here we consider a Mach-Zehnder interferometer with input state $|\psi(\tau)\rangle$
and infer the true value of the phase shift $\theta$ from the measurement of the 
relative number of particles at the output ports. 
These are characterized by conditional 
probabilities $P(\mu|j,\theta,\tau)=|_z\langle j, \mu| e^{-i \theta \hat{J}_y} |\psi(\tau)\rangle|^2$, 
being $\mu$ the result of a measurement.
To achieve $\Delta \theta_{\mathrm{QCR}}$, Eq.(\ref{cr}), we 
consider a Bayesian estimation scheme \cite{nota21, Pezze_2007}.
In Fig.(\ref{comp},b) we plot the results of numerical simulations
for $\tau=1/\sqrt{N}$ and $\theta=\pi/2$.
We show $\Delta \theta$ as a function of the total number of particles 
used in the estimation process $N_\mathrm{T} =Np$, being $p=p_{\mathrm{opt}}=20$ the optimal number 
of independent measurements. 
The circles are numerical results (minima in the inset of (\ref{comp},b)) and 
the line is $\Delta \theta = 8.9/N_\mathrm{T}$. 
We emphasize that 
the more popular phase estimation scheme based on 
the measurement of average moments of $\hat{J}_z$ \cite{Wineland_1992} and the corresponding
error propagation analysis only provide shot-noise.
\begin{figure}[!t]
\begin{center}
\includegraphics[scale=0.3]{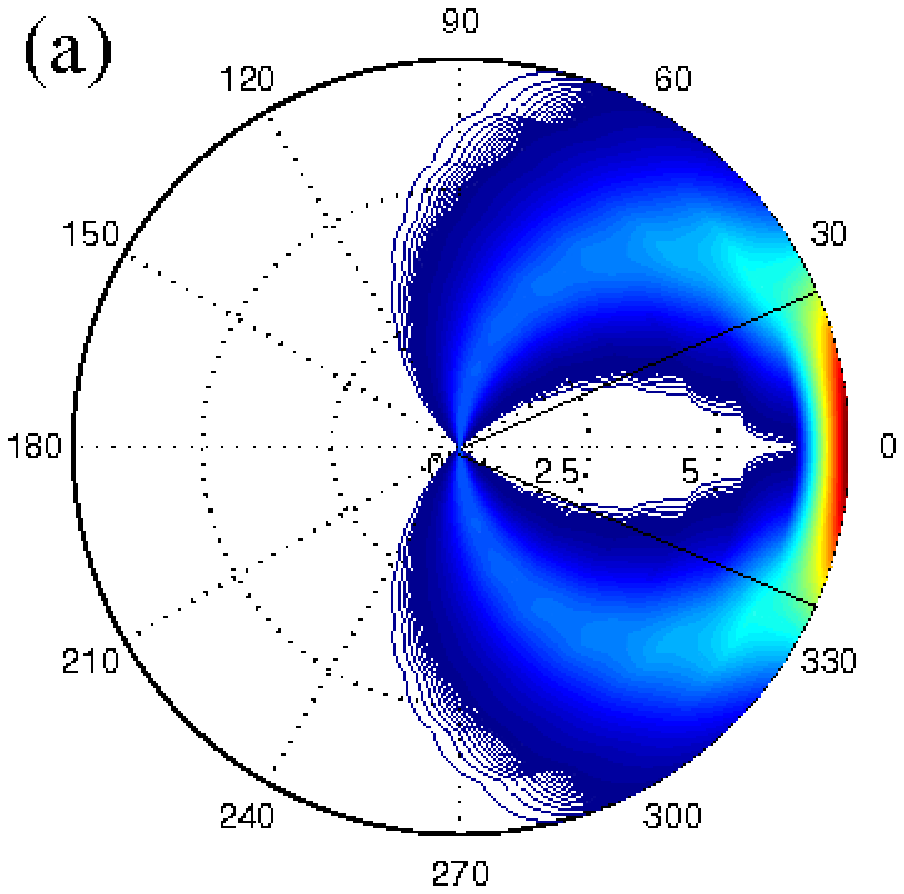}
\includegraphics[scale=0.18]{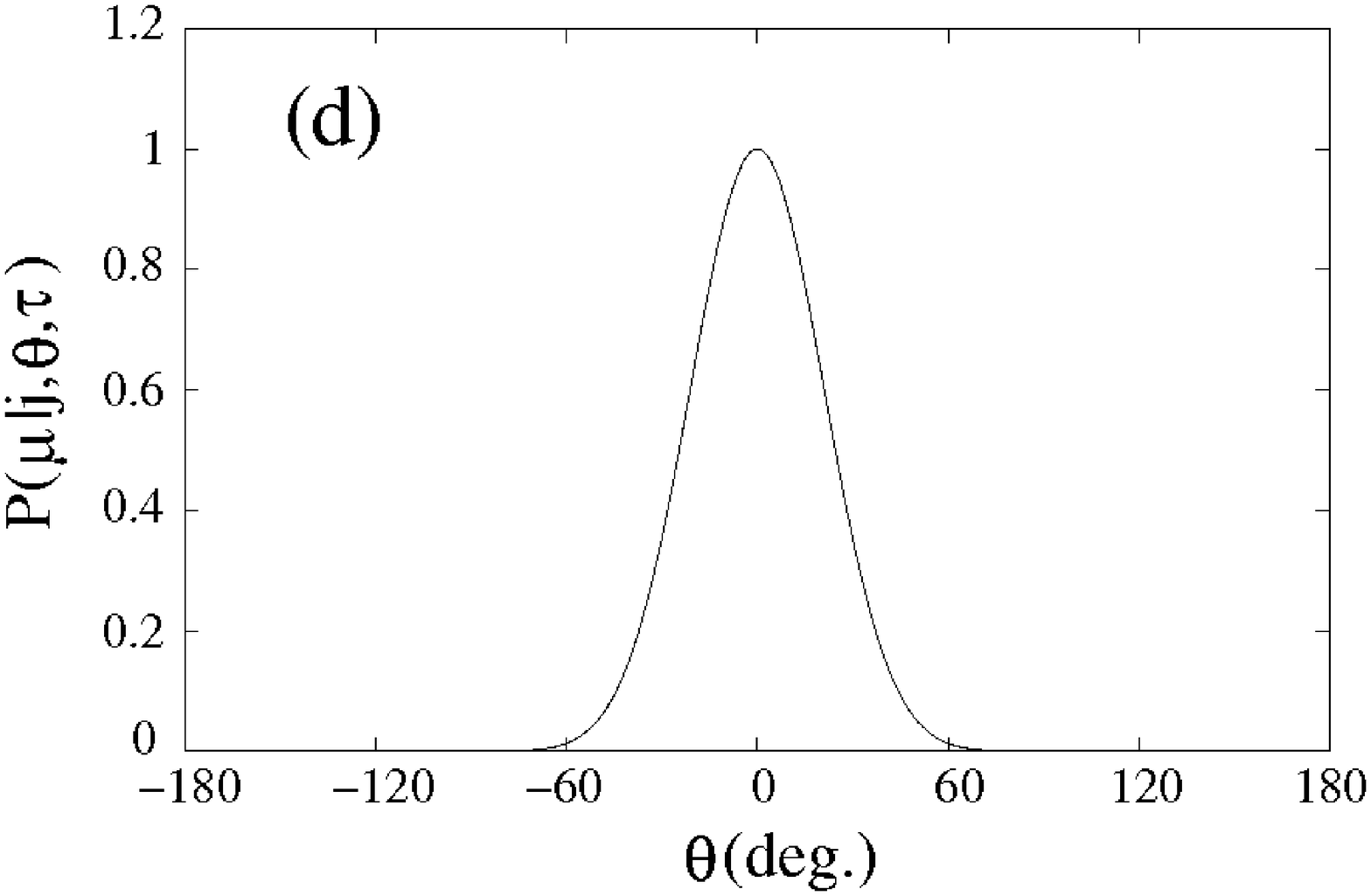}
\includegraphics[scale=0.145]{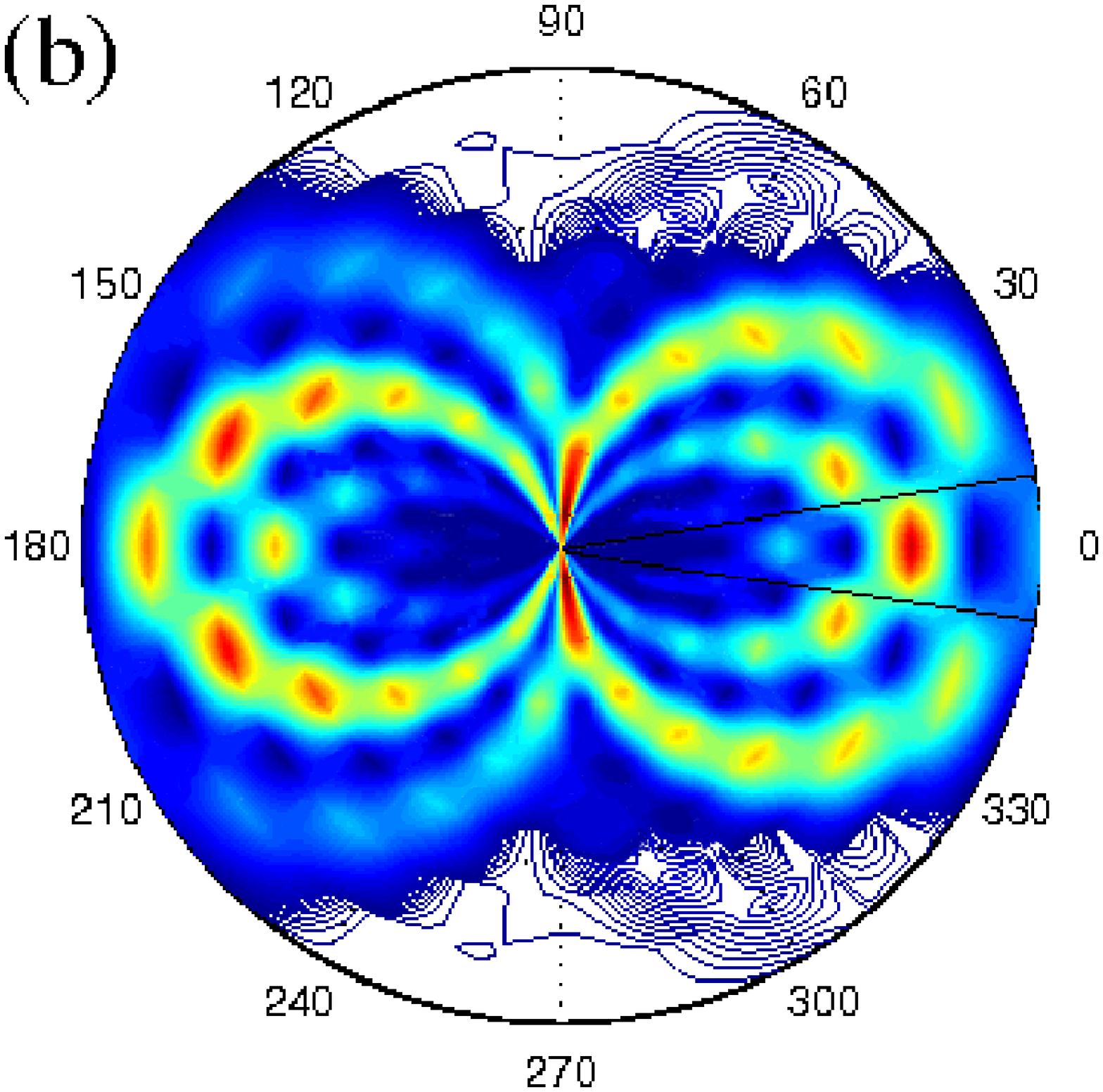}
\includegraphics[scale=0.18]{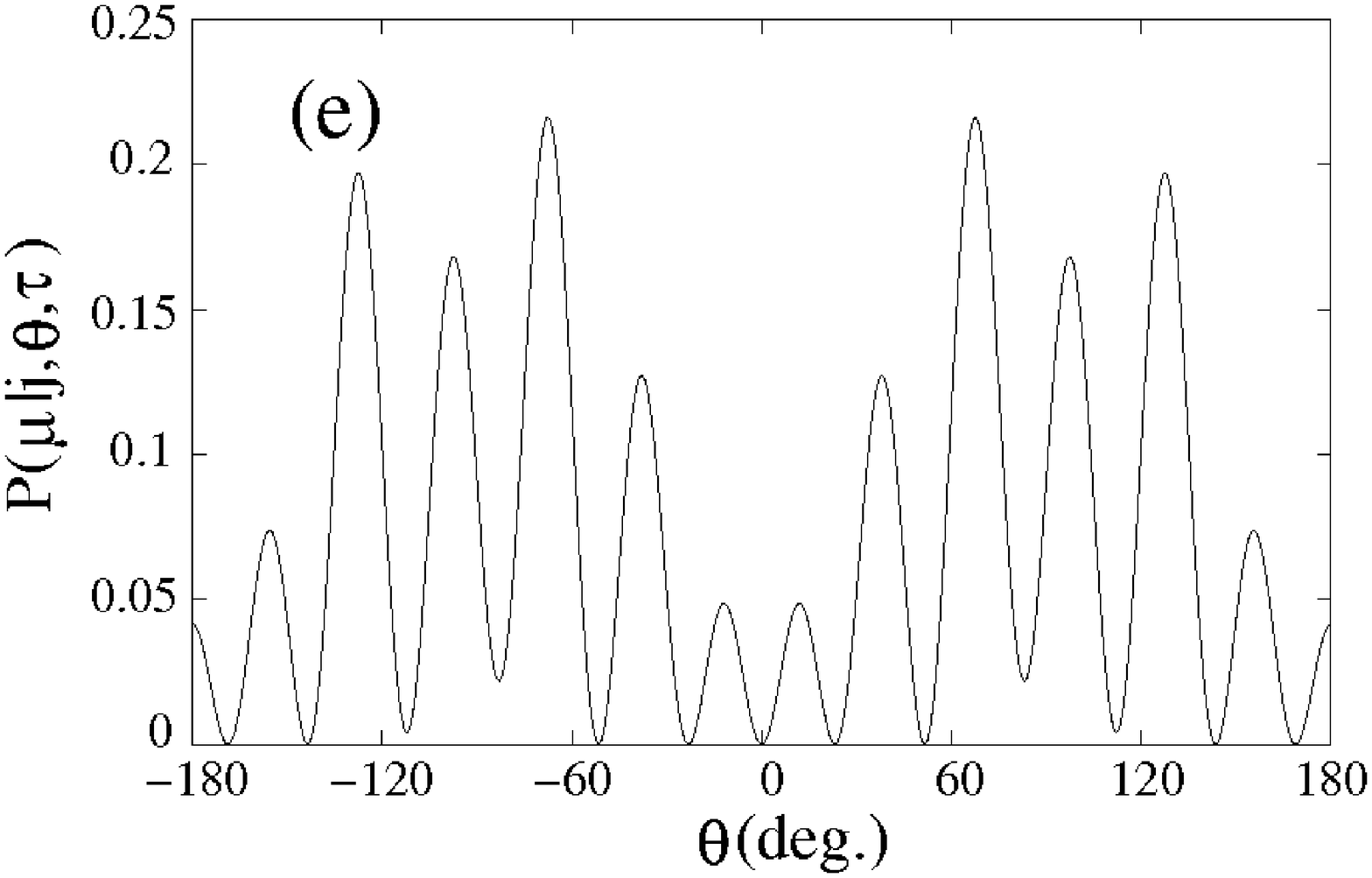}
\includegraphics[scale=0.3]{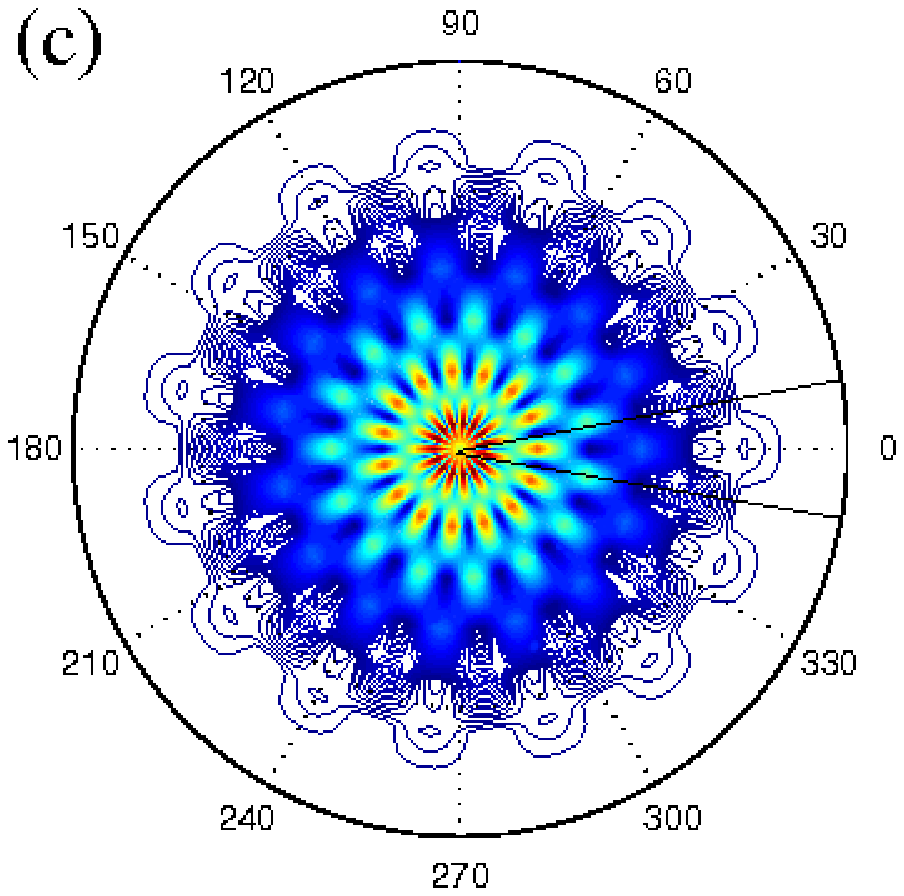}
\includegraphics[scale=0.18]{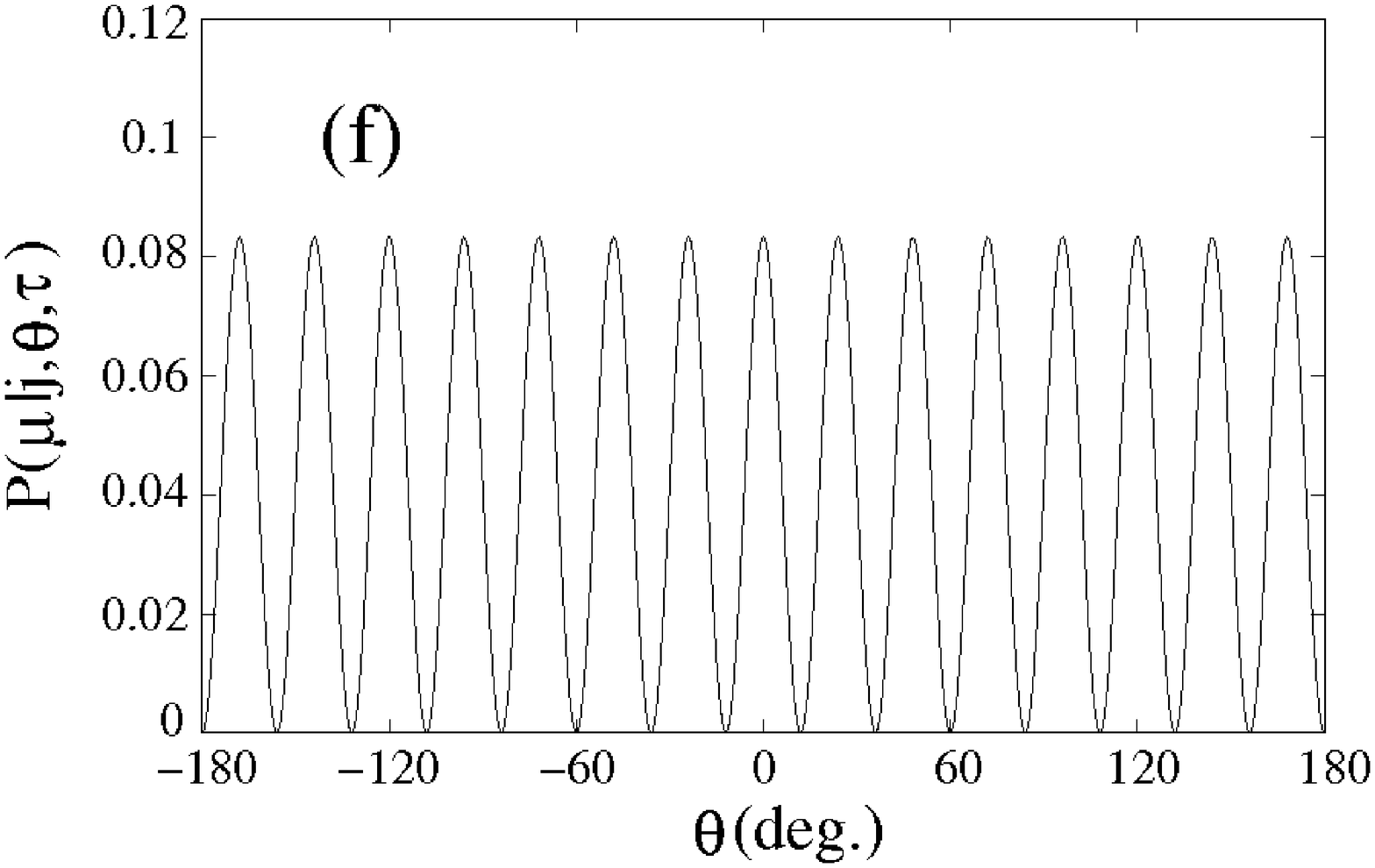}
\end{center}
\caption{\small{
(color online) a-c) Distributions $P(\mu|j, \theta, \tau)$
plotted as a function of $\theta$ along circles of radius $\mu$ 
(taken as a continuum variable) at three different times during the non-linear evolution:
a) $\tau=0$, b) $\tau=\pi/4$ and c) $\tau=\pi/2$. 
The solid lines delimit the typical size of the substructures. 
d-f) $P(\mu|j, \theta, \tau)$ as a function of $\theta$, 
and for: d) $\tau=0$, $\mu=7.5$ e) $\tau=\pi/4$, $\mu=2.5$ and f) $\tau=\pi/2$, $\mu=3.5$. Here $N=15$. }} \label{phasestructure} 
\end{figure}

Can we understand the origin of sub shot-noise without spin-squeezing ?
Let us investigate the phase structures characterizing the 
conditional probability distributions $P(\mu|j, \theta, \tau)$, defined for 
discrete values of $-j\leq \mu \leq j$.
These distributions contain all of the available information about the parameter $\theta$
that can be extracted from the measurement of $\mu$.
In Figs.(\ref{phasestructure},a)-(\ref{phasestructure},c) 
we plot 
$P(\mu|j, \theta, \tau)$,
as a function of $\theta$, along circles of radius $\mu$, 
at three different times during the non-linear evolution:
(\ref{phasestructure},a) $\tau=0$, 
(\ref{phasestructure},b) $\tau=\pi/4$ and 
(\ref{phasestructure},c) $\tau=\pi/2$.
The typical size of the substructures is $\sim 1/\sqrt{N}$ in 
(\ref{phasestructure},a) and $\sim 1/N$ in (\ref{phasestructure},b)
and (\ref{phasestructure},c) as indicated, in the figure, by solid lines.
This is also shown in Figs.(\ref{phasestructure},d)-(\ref{phasestructure},f)  
where we plot $P(\mu|j,\theta, \tau)$ for different $\mu$ and 
the same $\tau$ as in Figs.(\ref{phasestructure},a)-(\ref{phasestructure},c).
The size of the relevant substructures indicates 
the smallest rotation angle needed to make 
the rotated state orthogonal to the initial one.  
 
{\it Conclusion.}
We have explored the interplay between multiparticle entanglement and 
quantum interferometry.
A key role is played by the quantum Fisher information.
We obtained a sufficient condition for $N$-particles 
entanglement, $\chi < 1$, Eq.(\ref{ps}), which is more general than and incorporates
the spin-squeezing condition Eq.(\ref{cz}).
Large entanglement can be obtained through a non-linear evolution and 
used to reach 
a phase sensitivity at the Heisenberg limit.
Our results can have practical impact in precision spectroscopy, atomic clock and 
atomic/optical interferometry 
and can be implemented with Bose-Einstein condensates and trapped ions
within the current technology.


\begin{thebibliography}{99}
\bibitem{nota003} Eq.(\ref{ps}) calculated with 
$|\tilde{\psi}(\tau)\rangle$
gives $\chi^2=4/[4+(N-1)(A+\sqrt{A^2+B^2})]$ which, for $N\gg 1$, is
similar to Eq.(\ref{nosq}).
\bibitem{nota18} We refer to $\Delta \theta =\alpha/N$ as the
Heisenberg limit of phase sensitivity, provided, of course, that the prefactor 
$\alpha\geq 1$ does not depend on $N$.
\bibitem{nota21} We associate to the detection of a value $\mu$ the phase probability distribution 
$P(\phi|j, \mu)=P(\mu|j, \phi)P(\phi)/P(\mu)$ obtained from the Bayes theorem, 
where $P(\phi)$ is the prior knowledge (in the following $P(\phi)=1/\pi$) and 
$P(\mu)$ provides the normalization. 
The Bayesian distribution associated to $p$ independent measurements with results 
$\mu_1, ... , \mu_p$ is the product 
$P(\phi|j, \mu_1, ... , \mu_p) \sim \prod_{i=1}^{p}P(\phi|j, \mu_i)$.
The estimator $\theta_{\mathrm{est}}$ is chosen as the maximum of $P(\phi|j, \mu_1, ... , \mu_p)$, while 
phase sensitivity $\Delta \theta$ is the 68\% confidence interval around $\theta_{\mathrm{est}}$.
\end{thebibliography}
\end{document}